 \documentclass[final,5p,times,twocolumn,authoryear]{elsarticle}

\usepackage{amssymb}
\usepackage{lipsum}
\usepackage{amsmath}
\usepackage{algpseudocode}
\usepackage{subcaption}
\usepackage{setspace}
\usepackage{etoolbox}
\usepackage{graphicx}
\usepackage{float}
\usepackage{booktabs}
\usepackage{algorithm}

\usepackage{amsthm}

\journal{Future Generation Computer Systems}

\begin{document}

\begin{frontmatter}



\title{A Small Leak Sinks All: Exploring the Transferable Vulnerability of Source Code Models}


\author[scu]{Weiye Li}
\ead[scu]{wyli@stu.scu.edu.cn} 


\author[scu,cor1]{Wenyi Tang} 
\ead[scu]{wtang@scu.edu.cn}



\cortext[cor1]{Corresponding author.}

\affiliation[scu]{
    organization={Sichuan University},
    city={Chengdu},
    country={China}
}


\begin{abstract}
Source Code Model (SCM) aims to learn the proper embeddings from source codes, demonstrating significant success in various software engineering or security tasks. The recent explosive development of Large Language Model (LLM) extends the family of SCMs, bringing LLMs for code (LLM4Code) that revolutionize development workflows. Investigating different kinds of SCM vulnerability is the cornerstone for the security and trustworthiness of AI-powered software ecosystems, however, the fundamental one, transferable vulnerability, remains critically underexplored.
Existing studies neither offer practical ways, \textit{i.e.} require access to the downstream classifier of SCMs, to produce effective adversarial samples for adversarial defense, nor give heed to the widely used LLM4Code in modern software development platforms and cloud-based integrated development environments.
Therefore, this work systematically studies the intrinsic vulnerability transferability of both traditional SCMs and LLM4Code, and proposes a victim-agnostic approach to generate practical adversarial samples. 
We design a Hierarchical Adaptive Bandit-based Intelligent method for Transferable Attack (HABITAT), consisting of a tailored perturbation-inserting mechanism and a hierarchical Reinforcement Learning (RL) framework that adaptively selects optimal perturbations without requiring any access to the downstream classifier of SCMs. 
Furthermore, an intrinsic transferability analysis of SCM vulnerabilities is conducted, revealing the potential vulnerability correlation between traditional SCMs and LLM4Code, together with fundamental factors that govern the success rate of victim-agnostic transfer attacks.
These findings of SCM vulnerabilities underscore the critical focal points for developing robust defenses in the future.
Experimental evaluation demonstrates that our constructed adversarial examples crafted based on traditional SCMs achieve up to 64\% success rates against LLM4Code, representing over 15\% improvement over the existing state-of-the-art method.

\end{abstract}




\end{frontmatter}




\section{Introduction}
\label{introduction}
The advent of SCMs \cite{feng2020codebert,wang2021codet5,guo2022unixcoder,guo2020graphcodebert} has introduced powerful capabilities of software engineering and security research. These models, inspired by breakthroughs in natural language processing like BERT \cite{devlin2019bert}, aim to encode source codes into proper embeddings that capture their semantic and syntactic properties. 
In task-specific software applications, a SCM outpts its encoded embeddings to downstream classifiers, such as code classification \cite{zhang2019novel,alon2019code2vec}, code generation \cite{chen2021evaluating,wang2021codet5}, code repair \cite{chen2019sequencer,lutellier2020coconut}, clone detection \cite{svajlenko2014towards}, vulnerability detection \cite{zhou2019devign}, etc.
Traditional SCMs primarily focused on encoder-only and encoder-decoder architectures, demonstrating effectiveness in understanding and manipulating code for specific downstream tasks \cite{nijkamp2022codegen}. The emergence of LLMs has catalyzed a paradigm shift toward decoder-only architectures trained on large-scale text and code repositories such as GitHub \cite{chen2021evaluating}, expanding the scope from task-specific code understanding to general-purpose code generation and reasoning. This evolution has given rise to LLM4Code systems \cite{roziere2023code,lozhkov2024starcoder,guo2024deepseek,li2023starcoder}, which leverage massive pre-training to achieve unified capabilities across diverse coding scenarios \cite{wan2024deep}. 
The transition 
has dramatically expanded the scope and sophistication of code understanding and generation capabilities. Yet despite these remarkable advances, the security implications of this architectural evolution remain critically underexplored.  

One particular concern is the SCM's vulnerability to adversarial examples, for instance, minor syntactic changes (e.g., adding dead code, renaming variables) can cause a misclassification, thus exposing systemic risks in downstream software engineering applications.
The major body of current works has paid attention to the basic adversarial examples of SCM, but overlooked the transferable ones \cite{papernot2016transferability,liu2016delving}.
Considering the transfer attack scenario, a source SCM could be used to generate adversarial examples against a victim SCM and mislead the victim SCM's downstream classifier.
Recent research \cite{yang2024exploiting} has discovered that such transferable vulnerabilities exist among traditional SCMs, where adversarial examples crafted on one source SCM can transfer across multiple downstream classifiers.
Based on their reliance on victim classifier information, these transfer attacks fall into two categories: victim-aware approaches that leverage downstream model information, and victim-agnostic methods that operate without victim knowledge. This means that despite architectural differences across SCM systems, they share fundamental vulnerabilities that enable systematic compromise of the entire ecosystem, suggesting that structural diversity provides insufficient protection against powerful transfer attacks. In detail, similar to the above example, a maliciously crafted input that triggers incorrect code classification in one SCM could systematically evade downstream security classifiers across different platforms, including the critical transition from traditional encoder-only models to modern LLM4Code systems deployed in production, creating cascading security failures across entire software ecosystems \cite{zou2023universal}. The investigation of these transferable examples helps us better understand the fundamental limitations of current models and develop more robust training procedures across the entire SCM ecosystem.

Current adversarial attack methods against SCMs \cite{nguyen2023adversarial,yang2022natural,zhang2020generating} predominantly follow victim-aware paradigms that require prior knowledge of target architectures and model parameters. While these approaches can achieve high attack success rates on specific systems, their practical implementation is severely constrained by the difficulty of obtaining detailed victim information in real-world scenarios. This reliance on victim-specific knowledge makes these attacks hard to deploy systematically across the landscape of SCM systems, from traditional encoder-only models to modern LLM4Code architectures. Consequently, the limited transferability of existing victim-aware methods has failed to provide effective adversarial examples for developing robust defense strategies. Without practical attack vectors that can realistically threaten SCM systems, security researchers lack the necessary tools to evaluate and strengthen defenses against systematic vulnerabilities. 

As SCMs increasingly integrate into critical development workflows, understanding and mitigating these transferable vulnerabilities has become essential for maintaining software security. However, to our best knowledge, there is no study into the fundamental mechanisms governing vulnerability transferability between traditional SCMs and LLM4Code systems. The security community lacks comprehensive empirical frameworks that can predict and explain why certain adversarial perturbations successfully transfer across different architectural paradigms. Recent evidence suggests that vulnerability patterns exhibit unexpected cross-system transferability, yet the underlying factors driving this phenomenon remain poorly understood. This knowledge gap hampers the development of robust defenses and leaves critical vulnerabilities in production deployments inadequately characterized, highlighting the urgent need for systematic investigation into the empirical foundations of cross-system vulnerability transfer.

To bridge these gaps, we propose a victim-agnostic approach HABITAT for defense that achieves high transferability while assuming absolute zero access to downstream models and tasks. We present a novel hierarchical reinforcement learning framework that systematically generates powerful and practical adversarial examples through a multi-stage process guided by source SCMs encoders. Our approach begins by reformulating adversarial code modification as a hierarchical contextual bandit problem, where we strategically select insertion positions based on semantic salience at the global level, while choosing from six types of syntax-preserving transformations at the local level through position-specific bandits. To enhance cross-model generalization, we construct a transferable attack memory that records not only successful transformation strategies but also their contextual reward distributions from previous attacks. This memory serves as a foundation for our multi-model guidance mechanism, which employs a model-selection bandit to dynamically determine the most applicable attack strategies from previous models' experiences for each position in the current victim SCM, resulting in per-position attack plans tailored to the victim SCM's latent vulnerability surface. We then evaluate these adversarial examples which crafted solely through access to traditional SCMs, against LLM4Code while maintaining semantic equivalence. Our comprehensive evaluation spans multiple traditional SCMs, LLM4Code systems, and code classification tasks, achieving up to 64\% success
rates against LLM4Code and over 15\% improvement over exisiting state-of-the-art on traditional SCMs, demonstrating that our method effectively bridges the existing transferability gap between traditional SCMs and modern LLM4Code systems. Meanwhile, we introduced several dominant factors that may caused this transferability, including the insert positions, the different insertion types success rates, local via global attentions distribution, the code snippet's complexity, and its semantic comprehension disparities.


Therefore, our contributions can be summarized as follows:

\begin{itemize}
    \item To the best of our knowledge, this is the first systematic study about the transferable vulnerabilities of both traditional SCMs and LLM4Code.
    \item We proposed the HABITAT method, which combines a tailored perturbation-insertion mechanism with a hierarchical RL framework that adaptively selects optimal perturbations without requiring access to downstream classifiers of SCMs.
    \item We conducted an intrinsic transferability analysis of SCM vulnerabilities, revealing potential vulnerability correlations between traditional SCMs and LLM4Code, along with dominant factors governing the success rate of victim-agnostic transfer attacks.
    \item We evaluate the effectiveness of our proposed method, showing that our constructed adversarial samples transferred from traditional SCMs to the LLM4Code achieve up to 64\% success rate, and outperform the state-of-the-art methods on different evaluation metrics.
\end{itemize}

\section{Related Work}
\textbf{Traditional SCMs:} 
Traditional SCMs evolved from encoder-only and encoder-decoder architectures designed for code understanding. CodeBERT \cite{feng2020codebert} pioneered the adaptation of bidirectional encoders to code, while GraphCodeBERT \cite{guo2020graphcodebert} incorporated data flow information for enhanced semantic comprehension. UnixCoder \cite{guo2022unixcoder} unified multiple code tasks in a single pre-trained model, and CodeT5 \cite{wang2021codet5} implemented a transformer encoder-decoder architecture for both generation and understanding tasks. These models demonstrated effectiveness across applications including code summarization \cite{hu2018deep}, search \cite{gu2018deep}, and defect detection \cite{zhou2019devign}, while Code2vec \cite{alon2019code2vec} and ASTNN \cite{zhang2019novel} innovated with structural code representations.

\textbf{Evolution of LLM4Code:}
LLM4Code marked a shift from task-specific encoder architectures to generative decoder-only models with larger parameter and training data. Codex \cite{chen2021evaluating} demonstrated that large-scale pre-training on GitHub repositories enabled code synthesis from natural language. This evolution continued with CodeGen \cite{nijkamp2022codegen}, which investigated scaling properties of autoregressive models, while open-source contributions included StarCoder \cite{li2023starcoder, lozhkov2024starcoder}, trained on over 80 programming languages. DeepSeek-Coder \cite{guo2024deepseek} advanced the field by combining extensive pre-training with instruction tuning, and CodeLlama \cite{roziere2023code} adapted the Llama 2 \cite{touvron2023llama} architecture specifically for programming tasks, achieving state-of-the-art performance across numerous benchmarks \cite{wan2024deep}.

\textbf{Adversarial Attacks on Code Models:}
As SCMs entered development workflows, adversarial vulnerabilities became critical concerns. Zhang et al. \cite{zhang2020generating} pioneered gradient-based adversarial examples, while Yang et al. \cite{yang2022natural} developed natural adversarial examples preserving functionality. ALERT demonstrated variable renaming significantly impacts model predictions. Nguyen et al. \cite{nguyen2023adversarial} explored code completion vulnerabilities, and Ramakrishnan et al. \cite{ramakrishnan2022backdoors} investigated backdoor attacks. These span from white-box methods requiring complete model access to victim-agnostic techniques operating without target knowledge.

\textbf{Transfer Attacks:}
Transfer attacks allow crafting examples on accessible models that transfer to inaccessible targets. Originally studied in computer vision \cite{papernot2016transferability, liu2016delving}, code model transferability faces unique challenges from programming languages' structured nature. Yang et al. \cite{yang2024exploiting} discovered transferable vulnerabilities across different SCM architectures, indicating shared weaknesses. Yefet et al. \cite{yefet2020adversarial} identified cross-model transferable perturbations for code naming tasks. However, factors influencing transferability between traditional SCMs and LLM4Code remain underexplored—a critical gap as industry adopts decoder-only architectures.

\textbf{Defense Methods:}
Defending against code model attacks presents challenges due to code's discrete nature and functionality requirements. Zhou et al. \cite{zhou2019devign} enhanced adversarial training with code property graphs focusing on semantic features. Rabin et al. \cite{rabin2021generalizability} investigated code canonicalization to mitigate syntactic variations. Zou et al. \cite{zou2023universal} proposed structure-aware universal defenses. However, existing approaches target specific attack vectors rather than fundamental vulnerability patterns enabling cross-model transferability, highlighting the need for strategies addressing common weaknesses across architectures.




\section{Method}

\setlength{\abovecaptionskip}{4pt}
\setlength{\belowcaptionskip}{4pt}

\setlength{\intextsep}{6pt plus 2pt minus 2pt}

\subsection{Overview}

In this framework, we propose HABITAT, a hierarchical, encoder-guided adversarial attack framework designed to exploit the transferability across SCMs of varying architectures. Our method operates in four stages:

\textit{Position-Aware Transformation Discovery (PATD).}
We reformulate adversarial code modification as a two-level contextual bandit problem. At the global level, we select insertion positions based on semantic salience, while at the local level, we select from six types of syntax-preserving code transformations, guided by position-specific bandits.

\textit{Transferable Memory Construction.} For each successful attack, we record not only the best transformation strategies but also their contextual reward distributions, forming a transferable “attack memory” for later reuse.

\textit{Preference-Guided Strategy Adaptation (PGSA).} During PGSA execution, we first check whether there is a stored memory (preference) for each code snippet. If such a memory exists, we adapt it to guide the generation of the adversarial example. Otherwise, we revert to the PATD process.

\textit{Adaptive Multi-Model Memory Transfer (MMMT).} To enhance cross-
architecture generalization, we introduce a model-selection bandit that dynamically determines which previous model’s memory is most applicable at each position for current victim SCM. This results in a per-position attack plan tailored to the victim SCM’s latent vulnerability surface.


\subsection{Threat Scenario Formulation}
Our threat model formalizes adversarial capabilities within the context of victim-agnostic code models.

\subsubsection{Attack Objectives}
We focus on targeted adversarial perturbations against SCMs that satisfy the dual constraints of misclassification and semantic preservation:
\begin{equation}
C(x_{adv}) = y_{adv} \neq y_{true} \land S(x_{adv}) \equiv S(x)
\end{equation}
Where $C(\cdot)$ is the victim classifier, $S(\cdot)$ denotes semantic equivalence, and $y_{true}$ is the correct label. This formulation specifically targets security-critical contexts where misclassifying vulnerable code as secure presents substantial risks to downstream systems—precisely the scenario where robustness guarantees are most essential yet frequently under-evaluated in practice.

\subsubsection{Motivation and Adversary's Capability}
As defensive mechanisms for SCMs grow increasingly sophisticated and LLM4Code achieves tremendous progress in code understanding and generation, real-world adversaries now predominantly operate under limited knowledge conditions. Modern attackers rarely possess comprehensive information about victim models' parameters, architectures, or downstream tasks, creating a significant disconnect between theoretical transferability concepts and their actual implementation in adversarial campaigns targeting code models.

Our research bridges this gap by generating powerful and practical adversarial examples that effectively affect LLM4Code decisions using only traditional SCM encoders. We systematically exploit the inherent transferability between SCMs, revealing how structural and semantic patterns in code representations create persistent vulnerability surfaces across disparate model architectures and training domains. We impose the most restrictive constraints to date: our attacks require no test-set sampling, no model parameters, purely leveraging traditional SCM encoders' feature space guidance. By reformulating transfer-based attacks as a multi-hierarchical Contextual Bandit Problem, our framework achieves victim-agnostic transferability with no dependency on target model feedback and theoretical convergence guarantees, establishing a framework that accurately reflects real-world adversarial scenarios where attack opportunities remain abundant through strategically crafted transferable examples.





\subsection{Hierarchical MAB for Position-Aware Code Transformation Discovery}
Based on this threat model, we formulate PATD as a two-level decision problem, and design a hierarchical contextual bandit framework to select optimal insertion points and code transformations without feedback from the victim SCM.

Our architecture consists a two levels of MAB instances:

Global Tier: UCB1 algorithm explores optimal perturbation strategies across model architectures.

Local Tier: Position-sensitive MAB selects attack locations based on code semantic importance.

This hierarchy is formalized as follows:

Let $P = \{p_1, p_2, ..., p_m\}$ be the set of $m$ important positions in the code, and $T = \{t_1, t_2, ..., t_n\}$ be the set of $n$ available code transformations. For each position $p_i \in P$, we maintain a separate MAB instance $M_i$ that learns the effectiveness of each transformation $t_j \in T$ when applied at position $p_i$.

\subsubsection{MAB Principles}

Multi-armed bandits (MAB) is a simple but very powerful framework for algorithms that make decisions over time under uncertainty \cite{slivkins2019introduction}. Its core is to learn the value of each arm (code conversion operation) through feedback (attack succeed or not) and balance the relationship between choosing known high-value actions (exploiting) and trying new actions (exploring).

Our work based on the UCB1 (Upper Confidence Bound) algorithm \cite{kuleshov2014algorithms}, which decision depends on:
\begin{equation}
\label{ucb1standard}
UCB(a) = Q(a) + \alpha \cdot \sqrt{\frac{2 \cdot \ln(t)}{N(a)}}
\end{equation}
Where:
\begin{itemize}
    \item $Q(a)$: The estimated value of arm $a$ (exploitation term)
    \item $\alpha$: Exploration parameter that controls the degree of exploration
    \item $t$: Total number of pulls across all arms
    \item $N(a)$: Number of times arm $a$ has been pulled
\end{itemize}

This formula consists of two key components that reveal the essence of the UCB1 algorithm:
\begin{itemize}
    \item The first term ($Q(a)$) favors selecting actions that have performed best historically (exploitation)
    \item The second term ($\alpha \cdot \sqrt{\frac{2 \cdot \ln(t)}{N(a)}}$) encourages trying actions that have been selected less frequently (exploration)
\end{itemize}

As Table~\ref{tab:codetransformations} had shown, in this work, we introduced 6 different types of code insertion.

\begin{table*}[htb!]
\centering
\small
\caption{Code Transformations in Python}
\label{tab:codetransformations}
\begin{tabular}{l c}
\toprule
\textbf{Label} & \textbf{Content (in Python)} \\
\midrule
C1 & Insert unreachable code with random variables:\verb|if False: unused_var_1234 = 'hello world!'| \\
\midrule
C2 & Insert never-executed loop with random variables:\verb|while False: unused_var_5678 = 'unreachable'; break| \\
\midrule
C3 & Declare unused random-numbered variables: \verb|unused_var_3456 = False| \\
\midrule
C4 & Define uncalled function with random naming: \verb|def unused_func_7890(): return None| \\
\midrule
C5 & Insert non-substantive comments: \verb|# NOTE: This is a comment| \\
\midrule
C6 & Add empty string print statements: \verb|print("")| \\
\bottomrule
\end{tabular}
\end{table*}

\subsubsection{Position-specific MAB}
For each code snippet, We first select the important positions, which contains a two-stage analysis:

\begin{figure}
\centering
\includegraphics[scale=0.082]{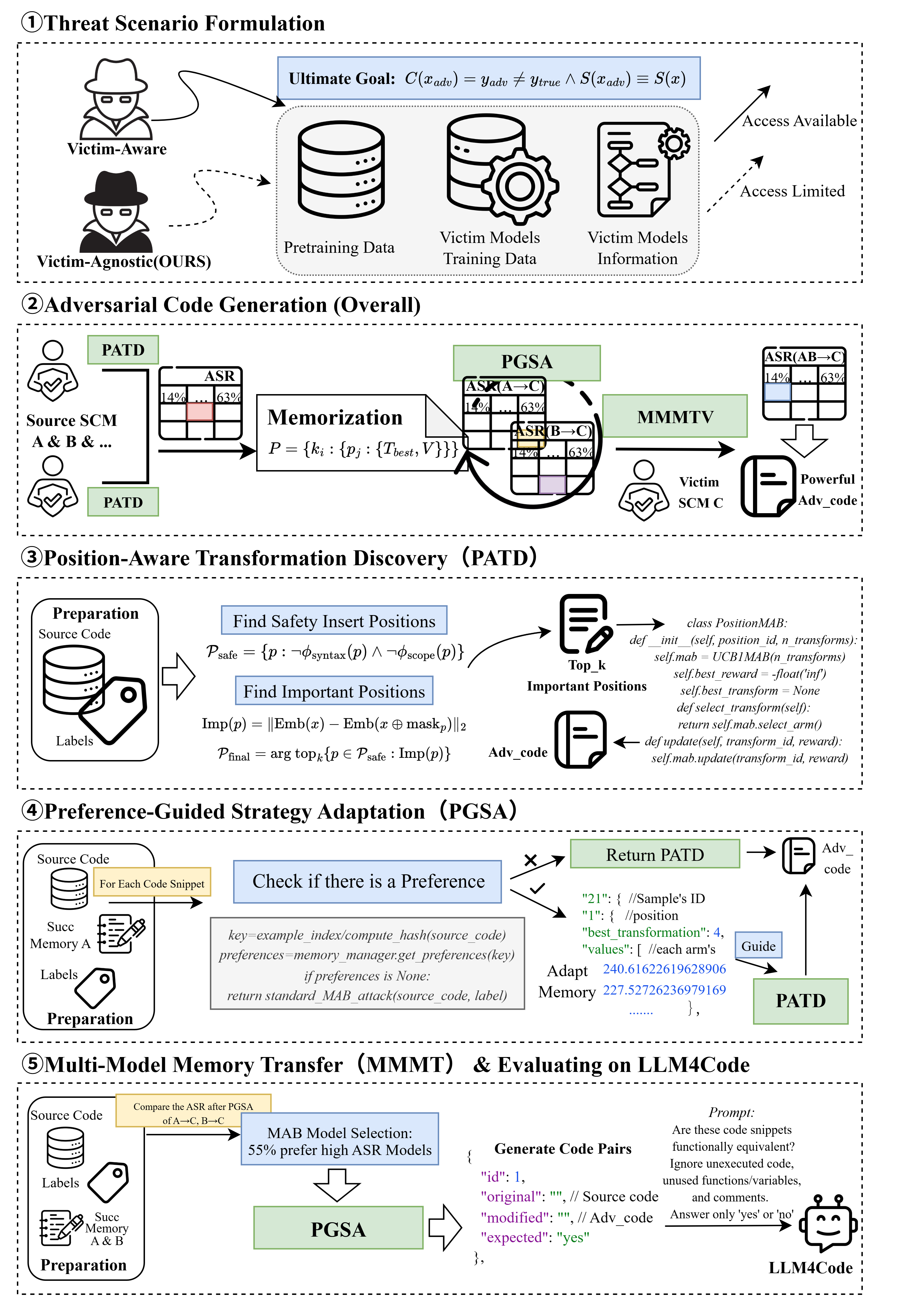}
\caption{HABITAT Pipeline}
\label{fig:transfer-attack-flow}
\end{figure}

\textbf{Safe Position Identification:} We first analyze the source code to identify syntactically safe insertion points that preserve program structure:
\begin{equation}
\mathcal{P}_{\text{safe}} = \{p : \neg\phi_{\text{syntax}}(p) \land \neg\phi_{\text{scope}}(p)\}
\end{equation}
where $\phi_{\text{syntax}}$ and $\phi_{\text{scope}}$ are predicates that detect syntactic violations and scope conflicts respectively. Our algorithm excludes:
\begin{itemize}
    \item Multi-line string and comment regions
    \item Import statements and function/class declarations
    \item Decorator lines and indentation-sensitive positions
\end{itemize}

\textbf{Semantic Importance Ranking:} From the safe positions, we select the top-$k$ most semantically important locations by computing feature-space impact:
\begin{equation}
\text{Importance}(p) = \|\text{Emb}(x) - \text{Emb}(x \oplus \text{mask}_p)\|_2
\label{equ:importancescore}
\end{equation}



Each position-specific MAB, denoted as $M_i$, encapsulates:
\begin{itemize}
    \item A unique position identifier $p_i$
    \item An internal UCB1 MAB instance with $n$ arms (one per transformation)
    \item State variables tracking the historically best-performing transformation and its associated reward
\end{itemize}

The position MAB manages the selection of transformations through a delegation pattern, as illustrated in equation~\ref{ucb1standard}.

\begin{algorithm}
\caption{Position-specific MAB}
\label{alg:position_mab}
\begin{algorithmic}[1]
\State \textbf{Initialize:} Position identifier $p_i$, number of transformations $n$
\State \textbf{Requires:} MAB object $M_i$, transformation ID $transformation\_id$, reward $reward$

\Procedure{Initialize}{$p_i$, $n$}
    \State $M_i.\text{position\_id} \gets p_i$
    \State $M_i.\text{mab} \gets \text{UCB1MAB}(n)$
    \State $M_i.\text{best\_transformation} \gets \text{null}$
    \State $M_i.\text{best\_reward} \gets -\infty$
\EndProcedure

\Procedure{SelectTransformation}{}
    \State \Return $M_i.\text{mab.select\_arm()}$
\EndProcedure

\Procedure{Update}{$transformation\_id$, $reward$}
    \State $M_i.\text{mab.update}(transformation\_id, reward)$
    \If{$reward > M_i.\text{best\_reward}$}
        \State $M_i.\text{best\_reward} \gets reward$
        \State $M_i.\text{best\_transformation} \gets transformation\_id$
    \EndIf
\EndProcedure
\end{algorithmic}
\end{algorithm}

${Critical}$  ${insight:}$
Our empirical analysis reveals that although there's a considerable transferability between different SCMs, the position importance distributions across different SCMs exhibit surprisingly low correlation. We will discuss this phenomenon particularly in the next section.



This observation, contrary to intuition, actually strengthens our approach: our hierarchical MAB architecture effectively learns position-specific strategies independently, making it robust to the absence of cross-architecture position correlations. 

The position-specific MAB overcomes this challenge by:
\begin{itemize}
    \item Learning optimal strategies independently for each position
    \item Not relying on pre-assumed position importance correlations
    \item Adapting to architecture-specific characteristics through online learning
\end{itemize}
\subsection{Memory-Augmented Mechanisms}
While the hierarchical bandit enables efficient attack on individual victim models, we further enhance transferability by storing transformation patterns and reward histories from successful attacks, which allowing us to reuse and adapt past strategies across samples.

\subsubsection{Persistent Memory Architecture for Attack Transferability}
For each successful attack, we store a comprehensive preference profile that captures both the optimal transformation strategies and their associated reward distributions:
\begin{equation}
    \mathcal{M}(x) = \{(p_i, t_i^*, \mathbf{V}_i) : \forall p_i \in \mathcal{P}_{\text{important}}\}
\end{equation}
where $p_i$ denotes the position index, $t_i^*$ represents the optimal transformation for position $p_i$, and $\mathbf{V}_i \in \mathbb{R}^{|\mathcal{T}|}$ encodes the empirical reward vector for all transformations at position $p_i$. The reward for each transformation is computed as:
\begin{equation}
    r_{t,p} = \|\text{Emb}(x) - \text{Emb}(x \oplus \mathcal{T}_t(p))\|_2 + [\text{Attack}_{\text{success}}(x \oplus \mathcal{T}_t(p))]
\end{equation}
This composite reward function captures both the semantic perturbation magnitude in the feature space and the actual attack efficacy, enabling nuanced transfer of attack strategies.

\begin{algorithm}
\caption{Adaptive Multi-Model Selection}
\label{alg:adap}
\begin{algorithmic}[1]
\State \textbf{Input:} Source code $S$, Model experiences $E_1, E_2$, Positions $P$
\State \textbf{Output:} Adversarial code $S'$
\State Initialize model selection MABs: $M_{\text{model}}[p] \gets \text{UCB1MAB}(2)$ for $p \in P$
\State Initialize transformation MABs using stored experiences $E_1, E_2$
\For{iteration $= 1$ to max\_iterations}
    \For{each position $p \in P$}
        \State model\_idx $\gets$ MAB selection (70\%) or random (30\%)
        \State transform\_idx $\gets$ Select transformation using chosen model's MAB
        \State Store selections: selected\_transforms[$p$], model\_choices[$p$]
    \EndFor
    \State $S_c \gets$ Apply transformations to $S$
    \State reward $\gets$ Evaluate attack effectiveness of $S_c$
    \State Update transformation and model selection MABs with reward
    \If{attack succeeds} \Return $S_c$ \EndIf
\EndFor
\State \Return best adversarial example found
\end{algorithmic}
\end{algorithm}
\subsection{Adaptive Multi-Model Memory Transfer}
To maximize attack effectiveness across different victim SCMs, we propose an Adaptive multi-model approach that intelligently leverages knowledge from multiple previous successful attacks. As illustrated in Figure~\ref{fig:transfer-attack-flow}, before MMMT, we first perform a comprehensive pairwise ASR (Attack Success Rate) analysis across source and victim SCMs. This step identifies which individual source SCM produces more transferable adversarial examples for a given target. These ASR values are stored and used as priors to inform the MMMT process, this method combines and adaptively selects from the attack preferences learned from different victim SCMs.
\subsubsection{Hierarchical Bandit Structure}
We also abstract the problem by using a hierarchical multi-armed bandit structure with two levels of decision-making:
\begin{enumerate}
    \item \textbf{Model Selection Level}: For victim model C and its each position, which model's attack experience provides better guidance? (which one shares more vulnerability with C?)
    \item \textbf{Transformation Selection Level}: For each code snippet, given the selected model's memory(if there is one), which transformation should be applied?
\end{enumerate}
\subsubsection{Adaptive Multi-Model Knowledge Selection}
Different victim models often exhibit distinct vulnerabilities at different code positions, making static or uniform knowledge transfer suboptimal. To address this, we propose an adaptive multi-model selection strategy that dynamically chooses which prior SCM’s experience to leverage at each important code location. Instead of applying a single SCM’s preferences globally or averaging across models, we maintain a position-specific MAB selector that learns, over time, which model offers the most effective guidance at each location. This enables fine-grained adaptation, allowing the attack to draw on Model A’s knowledge in some parts of the code and Model B’s in others, tailored to the victim SCM’s unique behavior. The details are presented in Algorithm~\ref{alg:adap}.
Our adaptive model selection process works as follows:
\begin{enumerate}
    \item For each position, we maintain a model-selection MAB (2 arms: model A, model B)
    \item When selecting transformations, we first use the model-selection MAB to decide which model's experience to follow
    \item The chosen model's transformation preferences then guide the selection of the specific transformation
    \item After applying transformations and evaluating results, we update both:
    \begin{itemize}
        \item The transformation MAB of the chosen model (lower-level decision)
        \item The model-selection MAB (higher-level decision)
    \end{itemize}
    \item As iterations progress, each position's model-selection MAB biases toward the model providing better guidance
\end{enumerate}

This approach balances exploration and exploitation by combining MAB-guided selection with randomness in early stages, then relying more heavily on learned preferences in later stages.

\section{Analysis of the Dominant Factors on Transferable Vulnerability}
Understanding the fundamental mechanisms behind adversarial transferability in SCMs represents a critical yet underexplored frontier in software security research. 

The transferability observed in SCMs stems fundamentally from their ability to learn and represent the intrinsic, underlying properties of source code that are common across different tasks and even different model architectures. Pre-training on large, diverse code corpora allows models to capture generalized features and structures of programming languages\cite{hussain2020deep}. Research indicates that despite variations in design and pre-training objectives, different SCMs are capable of learning similar fundamental code characteristics, including syntactic and semantic information\cite{han2021pre}.

In this section, we present the first comprehensive investigation of cross-architecture adversarial transferability in SCMs, systematically identifying and quantifying the underlying mechanisms rather than merely documenting the phenomenon.

\subsection{Insert Positions}
\begin{figure}[htb!]
\centering
\begin{minipage}{0.23\textwidth}
    \centering
    \includegraphics[width=0.68\textwidth]{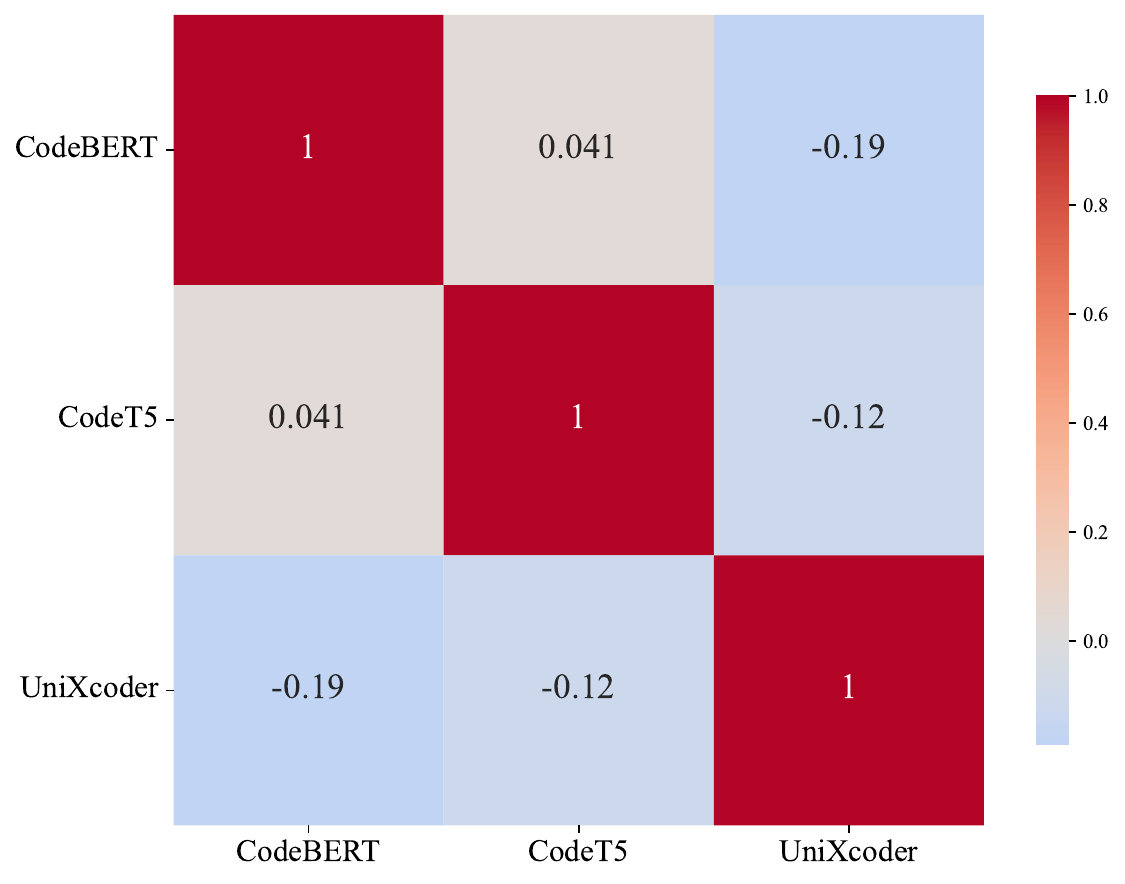}
    \caption{Samples position correlation}
    \label{fig:position-correlation}
\end{minipage}
\begin{minipage}{0.23\textwidth}
    \centering
    \includegraphics[width=0.74\textwidth]{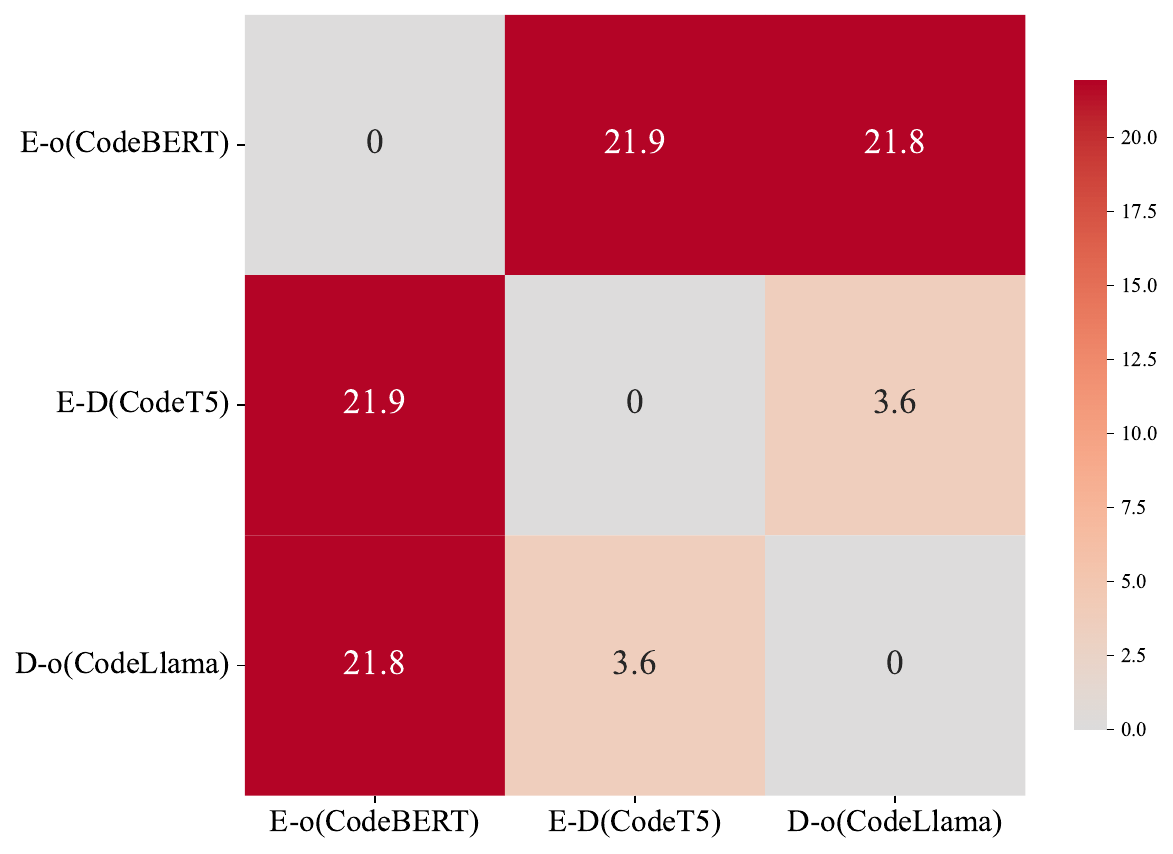}
    \caption{Model feature space similarity}
    \label{fig:Model Feature Space Similarity}
\end{minipage}
\end{figure}

We selected 100 code snippets from Authorship Attribution, as demonstrated in Figure~\ref{fig:position-correlation}, we quantify this through pairwise correlation analysis:
\begin{equation} 
\rho(M_i, M_j) = \text{Cov}(\mathbf{I}_{M_i}, \mathbf{I}_{M_j}) / (\sigma_{M_i} \sigma_{M_j}) 
\end{equation}
where $\mathbf{I}_M \in \mathbb{R}^{|\mathcal{P}|}$ represents the importance vector for model $M$ across all positions $\mathcal{P}$. Our experiments show that $\rho(M_i, M_j) < 0.1$ for most model pairs, especially between CodeBERT and UniXcoder, indicating that position sensitivity is model-specific rather than universal, therefore it is not a dominant factor that cause the cross-architecture transferability. 
\subsection{Success Rates of Different Insertions Across Different SCMs}
\begin{figure}[htb!]
\setlength{\belowcaptionskip}{-5pt} 
\centering
\includegraphics[scale=0.22]{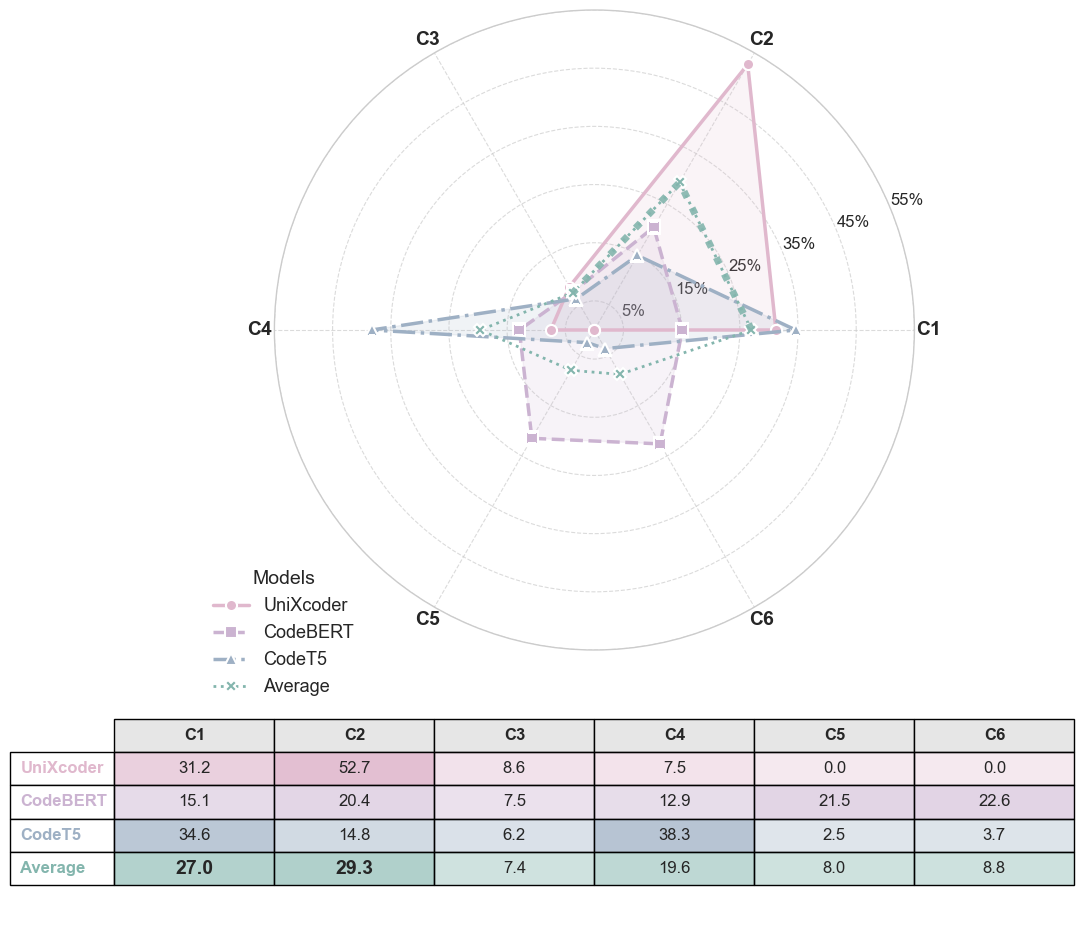}
\caption{Successful samples average insertion correlation}
\label{fig:transformtype-correlation}
\end{figure}

The statistical distributions in Figure~\ref{fig:transformtype-correlation} represent data from successfully attacked samples across different SCMs. Each model exhibits distinct vulnerability patterns to code transformation types (Table~\ref{tab:codetransformations}). C1 and C2 represent common vulnerabilities across all models with detailed differences: CodeBERT shows balanced vulnerability across six insertion types, UniXcoder exhibits strong bias toward C2 (52.7\%) while resisting C5-C6 transformations, and CodeT5 demonstrates particular weakness to C1 and C4 (34.6\% and 38.3\% respectively). These distinct vulnerability patterns explain their limitations across different programming contexts, enabling our approach to maximize transferability by leveraging each model's unique patterns rather than treating them as uniform systems.
While SCMs learn similar underlying code properties, they vary significantly in representation mastery \cite{ma2024unveiling}\cite{karmakar2021pre}. Different models show varying strengths in code syntax and semantics, with CodeT5 and CodeBERT excelling at capturing control flow and data flow dependencies compared to UniXcoder \cite{niu2023empirical}\cite{xiao2023empirical}. These differences stem from variations in architecture, pre-training objectives, and training data, leading to distinct internal representations and differential responses to adversarial perturbations. The existence of common vulnerability types (C1 and C2) provides transferability pathways, making insertion type a dominant factor in cross-architecture adversarial transferability.

\subsection{Local via Global Attentions Distribution}
\begin{figure}[htb!]
\setlength{\belowcaptionskip}{-5pt} 
\centering
\includegraphics[scale=0.18]{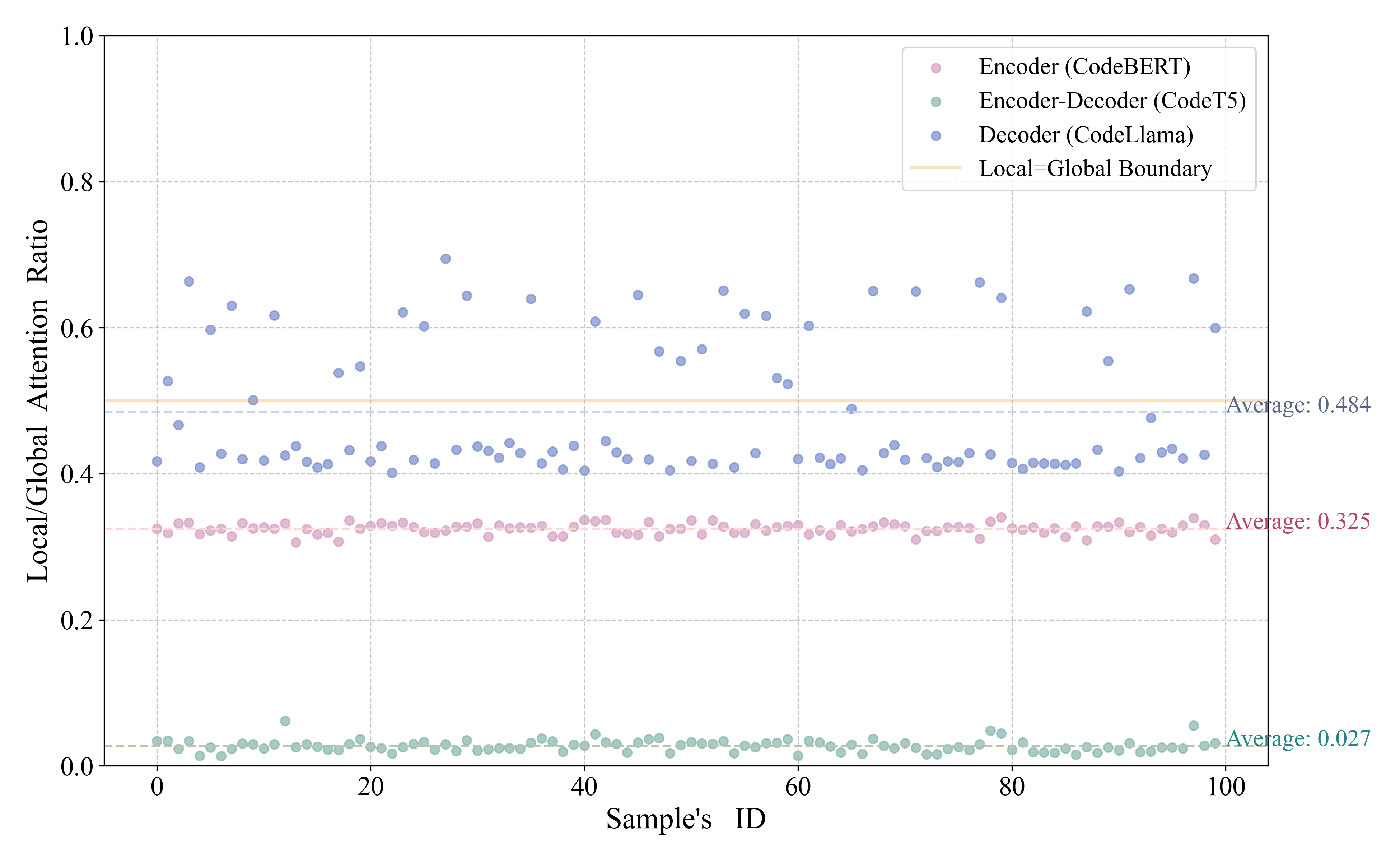}
\caption{Different SCMs Local/Global attention distribution}
\label{fig:attglobalvslocal}
\end{figure}
Attention mechanisms \cite{vaswani2017attention} play a crucial role in how different code models process and understand code structure. 

We quantified the Local/Global Attention Ratio ($\mathcal{R}_{LG}$) across 100 code samples from Authorship Attribution for three representative model architectures. For each model, we define the ratio as:
\begin{equation}
\mathcal{R}_{LG} = \frac{\sum_{i=1}^{n}\sum_{j=1}^{n} \mathbf{A}_{ij} \cdot \mathbf{1}(|i-j| \leq \delta)}{\sum_{i=1}^{n}\sum_{j=1}^{n} \mathbf{A}_{ij}}
\end{equation}
where $\mathbf{A}_{ij}$ represents the attention weight from token $i$ to token $j$, $n$ is the sequence length, $\delta=5$ is our local context window, and $\mathbf{1}(\cdot)$ is the indicator function that equals 1 when the condition is true and 0 otherwise.






Our results show significant architectural differences. Decoder-only models (CodeLlama) maintain the highest Local/Global Attention Ratio with an average of $\mathcal{R}_{LG} = 0.484$, indicating they allocate nearly equal attention to both local \& global code contexts. Encoder-only models (CodeBERT) show a moderate but consistent ratio averaging $\mathcal{R}_{LG} = 0.325$, while Encoder-Decoder models (CodeT5) exhibit a dramatically lower ratio of $\mathcal{R}_{LG} = 0.027$, indicating an overwhelming preference for global contextual understanding.


This analysis provides a direct explanation for the model-specific vulnerability patterns observed in our experiments and suggests that attention distribution characteristics represent a fundamental factor influencing adversarial transferability across different code model architectures.

\subsection{Length, Logical Structure and Complexity of Code Snippet}
Code complexity presents a multifaceted characteristic that potentially influences the cross-architecture transferability of adversarial examples. We systematically investigate this relationship through a multidimensional analysis of code structural properties and their impact on representation similarity across different architectures.

\subsubsection{Complexity Quantification Framework}
We define code complexity through a six-dimensional metric system, formalizing each dimension as follows:
\begin{enumerate}
    \item \textbf{Line Count} ($L_c$): Raw quantity of non-empty code lines
    
    \item \textbf{Character Count} ($C_c$): Total number of characters in the source code

    \item \textbf{Maximum Indentation Level} ($I_{max}$): Maximum nesting depth, where indentation level is calculated as:
    \begin{equation}
        I(line) = {\text{leading whitespace count}}/{4}
    \end{equation}

    \item \textbf{Control Structure Count} ($CS_c$): Frequency of control flow constructs:
    \begin{equation}
        CS_c = \sum_{line \in L_c} \mathbf{1}_{\exists k \in K: k \in line}
    \end{equation}
    {\small
    where $K = \{\text{if, else, elif, for, while, try, except, with, def, class}\}$
    }
    \item \textbf{Composite Complexity Score} ($C_{score}$): An aggregated metric synthesizing multiple dimensions:
    \begin{equation}
    \label{score}
        C_{score} = {L_c}/{10} + I_{max} + {CS_c}/{5}
    \end{equation}
\end{enumerate}

For comparative analysis, we categorize samples as ``High'' or ``Low'' complexity based on their position relative to the median $C_{score}$ across the dataset:
{\small
\begin{equation}
\text{Complexity Category}(s) = 
\begin{cases}
\text{High}, & \text{if } C_{score}(s) > \text{median}(C_{score}) \\
\text{Low}, & \text{otherwise}
\end{cases}
\end{equation}
}

\begin{figure}[htb!]
\centering
\includegraphics[scale = 0.18]{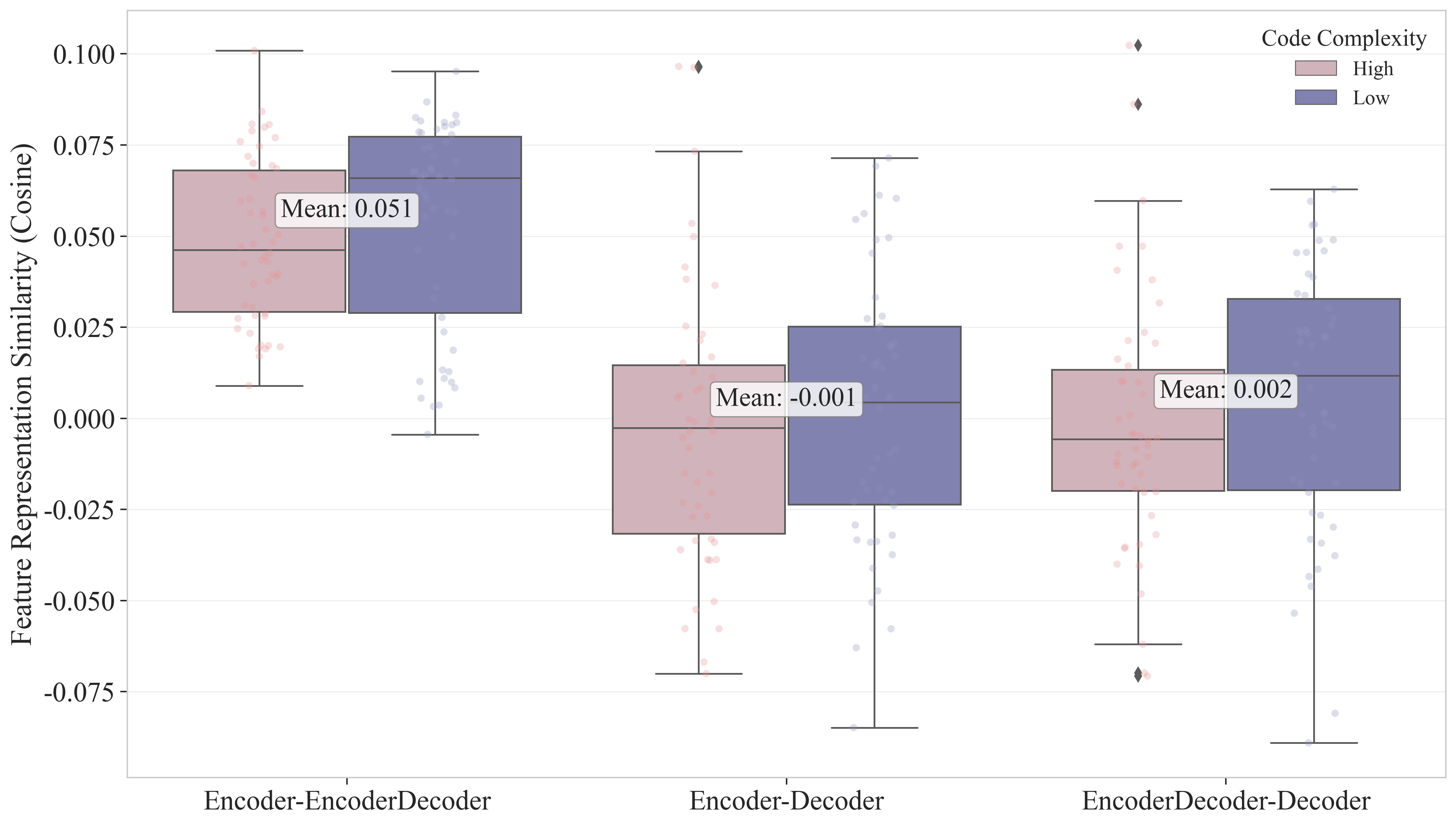}
\caption{Feature representation similarity across different model architecture pairs categorized by code complexity. }
\label{fig:model_pair_comparison}
\end{figure}

\subsubsection{Cross-Architecture Similarity Analysis}

For each architecture pair $(M_i, M_j)$, we calculate the cosine similarity between their feature representations:
\begin{equation}
\text{Sim}(M_i(s), M_j(s)) = {M_i(s) \cdot M_j(s)}/{||M_i(s)|| \cdot ||M_j(s)||}
\end{equation}

where $M_i(s)$ represents the feature vector generated by model $M_i$ for code sample $s$.

To investigate how code complexity affects cross-architecture transferability, we calculate separate mean similarity values ($\mu$) for high and low complexity subsets:

\begin{equation}
\mu_{M_i,M_j}^{High} = \frac{1}{|S_{High}|}\sum_{s \in S_{High}} \text{Sim}(M_i(s), M_j(s))
\end{equation}

\begin{equation}
\mu_{M_i,M_j}^{Low} = \frac{1}{|S_{Low}|}\sum_{s \in S_{Low}} \text{Sim}(M_i(s), M_j(s))
\end{equation}

where $S_{High} = \{s : \text{Complexity Category}(s) = \text{High}\}$ and $S_{Low} = \{s : \text{Complexity Category}(s) = \text{Low}\}$ are the high and low complexity sample sets respectively, determined by Equation (5). The overall mean similarity is:

\begin{equation}
\mu_{M_i,M_j} = \frac{|S_{High}| \cdot \mu_{M_i,M_j}^{High} + |S_{Low}| \cdot \mu_{M_i,M_j}^{Low}}{|S_{High}| + |S_{Low}|}
\end{equation}

This stratified analysis allows us to quantify whether code complexity systematically influences the consistency of feature representations across different model architectures.

\subsubsection{Results and Analysis}
Each data point in Figure~\ref{fig:model_pair_comparison} represents a code sample, with similarity values computed via Equation \ref{score}. The figure reveals key patterns in how code complexity affects cross-architecture feature similarity:

\begin{enumerate}
    \item \textbf{Architecture-Dependent Similarity}: Encoder-Encoder
    Decoder pairs show substantially higher similarity ($\mu=0.051$) than other combinations ($\mu \leq 0.002$), indicating architectural compatibility is the primary factor in representation similarity.
    
    \item \textbf{Variable Complexity Effects}: Code complexity impacts similarity differently across architecture pairs, with Encoder-EncoderDecoder favoring low-complexity code while Encoder-Decoder shows opposite trends.
    
    \item \textbf{Structural Outliers}: All pairs exhibit outlier samples with exceptional similarity values, indicating specific code structures maintain consistent representations across architectures regardless of complexity.
\end{enumerate}

While previous work~\cite{pierazzi2020intriguing} emphasized complex and redundant code for bypassing detection, our findings suggest that specific code structures with consistent cross-architecture representations are more effective than general code complexity. Although complexity is not the dominant factor in adversarial transferability, high-similarity outliers across architecture pairs indicate that certain samples can achieve notable cross-model transferability, with architectural compatibility as the primary determinant and specific structural patterns as secondary but significant factors.

\subsection{Semantic Comprehension Disparities}

To further understand the underlying mechanisms driving adversarial transferability across different model architectures, we analyze the feature distance between models using cosine similarity metrics. Figure~\ref{fig:Model Feature Space Similarity} presents a comprehensive distance matrix computed in our three above-mentioned target models.

The distance analysis reveals distinct clustering patterns among different architectural paradigms. Notably, CodeT5 and CodeLlama demonstrate the smallest inter-model distance (3.559), suggesting that these models operate in relatively similar feature spaces despite their architectural differences. This proximity in the feature space partially explains the higher transferability rates observed between encoder-decoder and decoder-only models in our empirical evaluations. This alignment can be theoretically supported by Fu et al. \cite{fu2023decoder}, who demonstrate that decoder-only language models can be interpreted as regularized encoder-decoder models, where the decoder component shares many functional similarities in how they process and generate sequences.

The distance values are computed using the Euclidean distance between feature vectors extracted from each model on a shared set of code samples:
\begin{equation}
D(M_i, M_j) = \frac{1}{|S|} \sum_{s \in S} \sqrt{\sum_{k=1}^{\min(|\mathbf{f}{i,s}|, |\mathbf{f}{j,s}|)} (f_{i,s,k} - f_{j,s,k})^2}
\label{eq:model_distance}
\end{equation}
where $\mathbf{f}_{i,s}$ represents the feature vector of sample $s$ in model $M_i$, and $S$ is the set of common samples across all models.

In contrast, CodeBERT exhibits significantly larger distances from both CodeT5 (21.945) and CodeLlama (21.846), with relatively uniform distances suggesting that the architectural gap between encoder-only and generation-capable models represents a more significant barrier to transferability than differences between encoder-decoder and decoder-only designs. These metrics provide quantitative evidence that models with similar training objectives and architectural components are more susceptible to shared vulnerabilities, facilitating adversarial transferability across model boundaries.


\section{Evaluation}
\subsection{Victim SCMs Training}

    In this study, we employ three state-of-the-art traditional SCMs as victim models: CodeBERT \cite{feng2020codebert}, CodeT5 \cite{wang2021codet5}, UniXcoder \cite{guo2022unixcoder}. We selected two code classification datasets, the Authorship Attribution\cite{alsulami2017source} and the Defeat Detection\cite{devlin2019bert}, the former originates from the Google Code Jam challenge and aims to identify the author of a given code snippet, while the latter is a C dataset that combines two popular open-sourced projects: FFmpeg3 and Qemu4.

Table \ref{tab:example} shows the training accuracy of our victim SCMs on clean datasets before adversarial attacks.

\begin{table}[H]
\caption{Clean Accuracy Comparison of Different Models on Various Datasets}
\label{tab:example}
\centering
\small
\begin{tabular}{ccc}
\toprule[0.8pt]
Dataset & Model & Clean Acc \\
\midrule[0.5pt]
& CodeBERT & 74.24 \\
\multicolumn{1}{c}{Authorship Attribution} & CodeT5 & 77.27 \\
& UniXCoder & 78.03 \\
\midrule[0.5pt]
& CodeBERT & 57.54 \\
\multicolumn{1}{c}{Defeat Detection} & CodeT5 & 58.97 \\
& UniXCoder & 44.14 \\
\bottomrule[0.8pt]
\end{tabular}
\end{table}

\subsection{Important Top\_k Position Chosen}
To determine the optimal number of insertion positions for PATD, we conduct a position selection experiment across different values of $k$ to balance attack effectiveness with code perturbation constraints.

\subsubsection{Experimental Setup}
We evaluate the impact of varying $k \in \{1, 2, 3, 5, 8, 10, 15\}$ across CodeBERT, CodeT5, and UniXcoder using three key metrics:
\textbf{Attack Success Rate (ASR):} Proportion of samples whose predictions are successfully changed.
\textbf{Feature Distance (FD):} Semantic distance between original and adversarial code representations, equivalent to equation~\ref{equ:importancescore}.
\textbf{Code Modification Rate (CMR):} Percentage of code tokens modified during attack.

\subsubsection{Position Selection Strategy}
\begin{table}[htbp]
\centering
\small 
\caption{ASR Comparison of Different Models with PATD and Baselines on Authorship Attribution}
\label{tab:model_performanceaa}
\begin{tabular}{lccccc}
\toprule[0.8pt]
Model & CodeTAE & PATD(ours) & Random-only(ours) \\
\midrule[0.5pt]
CodeBERT & 11.24(+14.29) & 25.53 & 15.96(+9.57) \\
\midrule[0.5pt]
CodeT5 & 5.88(+16.67) & 22.55 & 15.00(+7.55) \\
\midrule[0.5pt]
UniXcoder & 15.38(+14.29) & 29.67 & 19.78(+9.89) \\
\midrule[0.5pt]
Avg & 10.83(+15.09) & 25.92 & 16.88(+9.04) \\
\bottomrule[0.8pt]
\end{tabular}
\end{table}

\begin{table}[htbp]
\centering
\small 
\caption{ASR Comparison of Different Models with PATD and Baselines on Defeat Detection}
\label{tab:model_performancede}
\begin{tabular}{lccccc}
\toprule[0.8pt]
Model & CodeTAE & PATD(ours) & Random-only(ours) \\
\midrule[0.5pt]
CodeBERT & 5.86(+8.6) & 10.46 & 3.35(+7.11) \\
\midrule[0.5pt]
CodeT5 & 17.65(+22.99) & 40.64 & 19.25(+21.39) \\
\midrule[0.5pt]
UniXcoder & 8.23(+11.39) & 19.62 & 12.03(+7.59) \\
\midrule[0.5pt]
Avg & 10.58(+12.99) & 23.57 & 11.54(+12.03) \\
\bottomrule[0.8pt]
\end{tabular}
\end{table}

We evaluate PATD from two perspectives: attack effectiveness and stealthiness. Higher $k$ values improve ASR but increase Code Modification Rate and Feature Distance, compromising stealthiness and increasing detection risk. Therefore, we seek optimal $k$ that maximizes ASR while minimizing code modifications and semantic distance.

\subsubsection{Results and Analysis}
We select $k = 3$ as the optimal configuration for both datasets based on the trade-off between attack effectiveness and stealthiness. Detailed ASR, FD and CMR results for all tested $k$ configurations are presented in Table~\ref{tab:model_performanceaa} and
~\ref{tab:model_performance_defeat}.

\begin{table}[H]
\caption{SCM FD \& CMR \& ASR [\%] Comparison Across Different Top-k Positions on Authorship Attribution}
\label{tab:model_performanceaa}
\centering
\scriptsize 
\begin{tabular}{ccccccccc}
\toprule[0.8pt]
Models & k=1 & k=2 & k=3 & k=5 & k=8 & k=10 & k=15  \\
\midrule[0.5pt]
 & 217.1 & 218.6 & 221.4 & 226.7 & 224.3 & 225.6 & 226.5 \\
 CodeBERT & 9.9 & 18.9 & 28.3 & 48.7 & 74.4 & 88.9 & 113.0 \\
 & 24.47 & 30.85 & 32.98 & 32.98 & 36.17 & 37.23 & 41.49\\
\midrule[0.5pt]
& 130.3 & 154.9 & 141.7 & 179.3 & 253.5 & 375.9 & 1591.8 \\
CodeT5 & 7.9 & 15.5 & 22.1 & 39.2 & 59.5 & 75.1 & 104.7  \\
 & 24.51 & 24.51 & 24.51 & 29.41 & 39.22 & 46.08 & 57.84 \\
\midrule[0.5pt]
 & 352.0 & 480.6 & 457.9 & 486.3 & 509.0 & 468.0 & 502.5 \\
 UniXcoder & 26.0 & 46.7 & 64.3 & 89.7 & 98.8 & 98.7 & 99.3  \\
 & 20.88 & 29.67 & 29.67 & 40.66 & 50.55 & 51.65 & 50.55 \\
\bottomrule[0.8pt]
\end{tabular}
\end{table}

\begin{table}[H]
\caption{SCM FD \& CMR [\%] \& ASR [\%] Comparison Across Different Top-k Positions on Defeat Detection}
\label{tab:model_performance_defeat}
\centering
\scriptsize 
\begin{tabular}{ccccccccc}
\toprule[0.8pt]
Models & k=1 & k=2 & k=3 & k=5 & k=8 & k=10 & k=15  \\
\midrule[0.5pt]
 & 1.31 & 1.74 & 2.07 & 2.56 & 3.00 & 3.17 & 3.42 \\
CodeBERT & 2.5 & 4.4 & 6.3 & 9.4 & 13.4 & 15.5 & 19.3 \\
 & 8.37 & 7.11 & 11.72 & 11.30 & 14.64 & 15.48 & 15.90 \\
\midrule[0.5pt]
 & 0.85 & 1.10 & 1.37 & 1.63 & 2.08 & 2.05 & 2.64 \\
CodeT5 & 1.9 & 3.3 & 4.9 & 6.9 & 10.0 & 11.2 & 15.9 \\
 & 36.90 & 39.57 & 41.48 & 45.45 & 47.06 & 49.73 & 44.92 \\
\midrule[0.5pt]
 & 5.17 & 6.57 & 7.68 & 10.03 & 12.34 & 13.54 & 15.21 \\
UniXcoder & 2.3 & 4.0 & 5.6 & 8.8 & 12.3 & 14.3 & 17.5 \\
 & 13.29 & 18.35 & 21.52 & 19.62 & 24.05 & 23.42 & 24.68 \\
\bottomrule[0.8pt]
\end{tabular}
\end{table}

\subsection{Attack Effectiveness Across Exploration Rates}
We design experiments that systematically explore the trade-off between exploration and exploitation in multi-model scenarios. 

\begin{figure*}[htbp]
   \centering
   \begin{subfigure}[t]{0.33\textwidth}
       \centering
       \raisebox{-0.033cm}{\includegraphics[height=3.2cm]{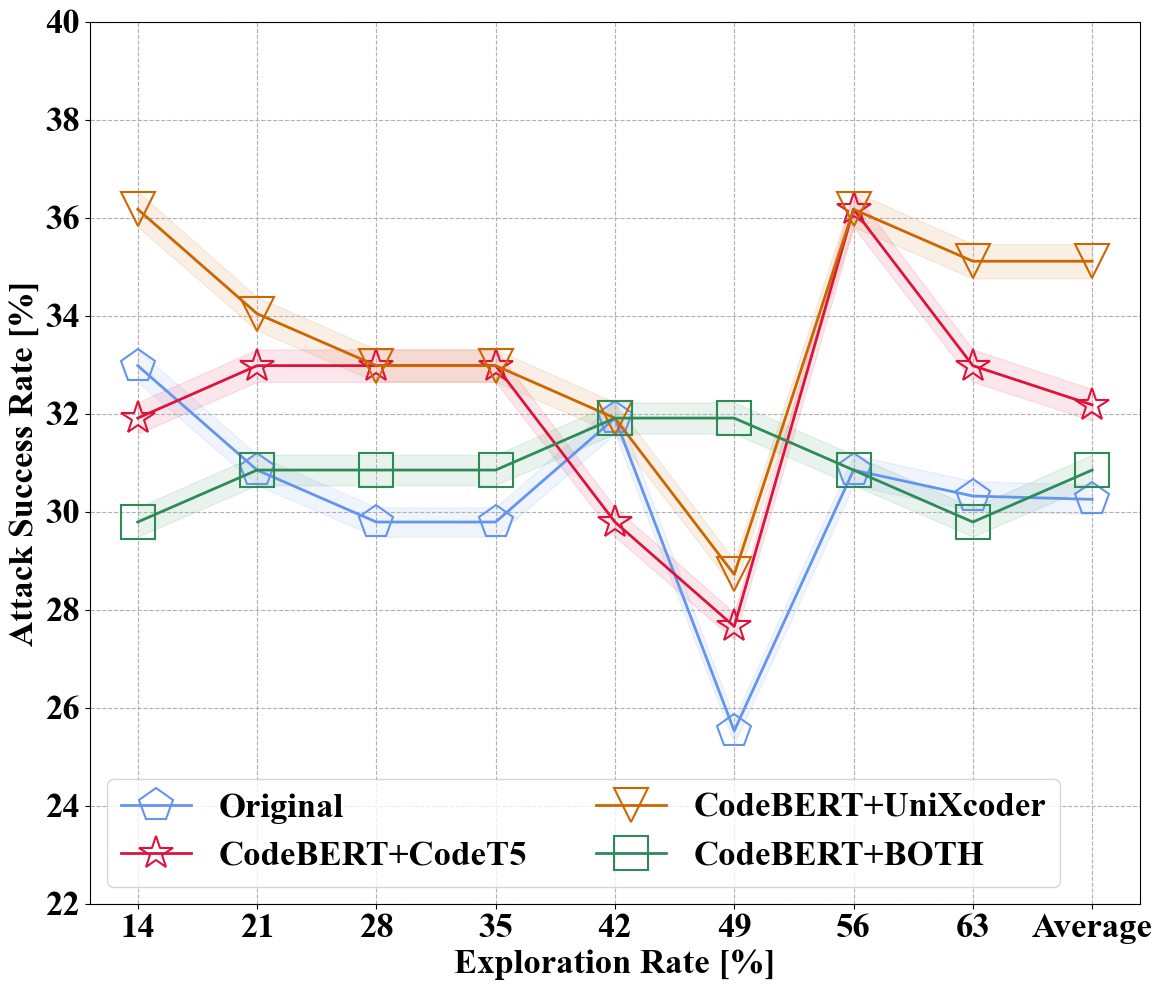}}
       \caption{ASR Comparison of CodeBERT through PATD, PGSA, MMMT}
       \label{fig:codebert}
   \end{subfigure}
   \hfill
   \begin{subfigure}[t]{0.33\textwidth}
       \centering
       \raisebox{-0.04cm}{\includegraphics[height=3.2cm]{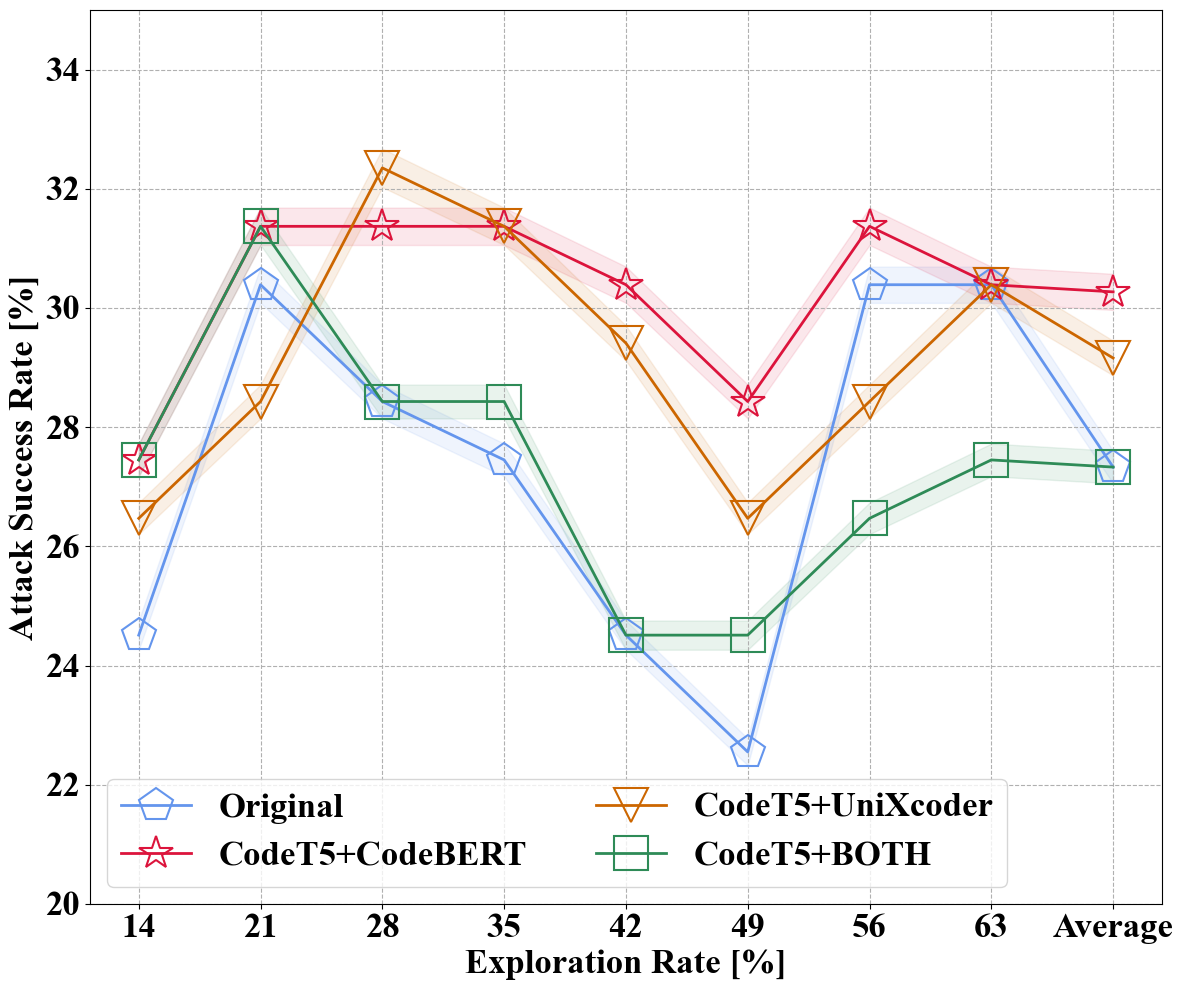}}
       \caption{ASR Comparison of CodeT5 through PATD, PGSA, MMMT}
       \label{fig:codet5}
   \end{subfigure}
   \hfill
   \begin{subfigure}[t]{0.33\textwidth}
       \centering
       \raisebox{-0.052cm}{\includegraphics[height=3.28cm]{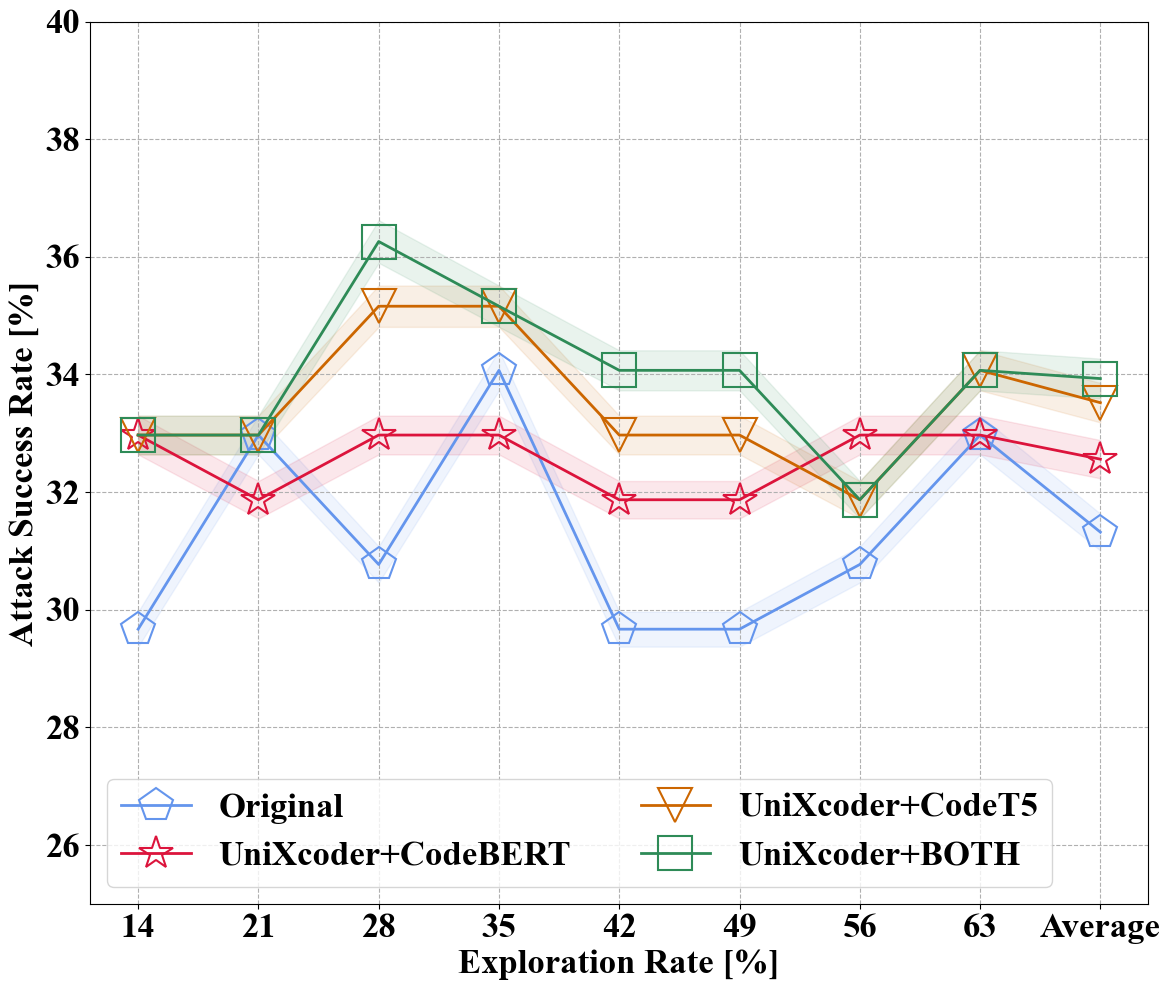}}
       \caption{ASR Comparison of UniXcoder through PATD, PGSA, MMMT}
       \label{fig:unixcoder}
   \end{subfigure}
   \caption{Comparison of different traditional SCMs on Authorship Attribution}
   \label{fig:threemodels}
\end{figure*}

\subsubsection{Baseline}
We adopt CodeTAE \cite{yang2024exploiting} as our baseline, and we also designed a random-only baseline for the PATD method that uses only random combinations without the MAB exploration-exploitation process.

\subsubsection{Evaluation Metric}
The effectiveness of the attack is quantified by the \textbf{Attack Success Rate (ASR)}, defined as the fraction of initially correctly classified samples that are misclassified after the attack:
\[
\text{ASR} = \frac{|\mathcal{S}_{\text{success}}|}{|\mathcal{S}_{\text{correct}}|}
\]
where $\mathcal{S}_{\text{correct}}$ denotes the set of samples correctly classified by the victim model prior to the attack, and $\mathcal{S}_{\text{success}} \subseteq \mathcal{S}_{\text{correct}}$ is the subset that is misclassified post-attack.

\subsubsection{Experimental Parameters}

We systematically vary the exploration rate from 14\% to 63\% in 7\% increments to understand how guidance levels influence attack effectiveness. This rate equals MAB exploration rate × 0.7, with 70\% MAB-directed exploitation and 30\% random exploration. Lower rates emphasize exploration, while higher rates focus on exploitation.

\begin{table}[htb!]
\centering
\scriptsize
\caption{ASR (\%) Results of CodeBERT with Different Exploration Rates and Strategies on Defeat Detection}
\setlength{\tabcolsep}{3pt}
\label{tab:codebert_results}
\begin{tabular}{lcccccccccc}
\toprule[0.8pt]
Strategy & 14\% & 21\% & 28\% & 35\% & 42\% & 49\% & 56\% & 63\% & Avg \\
\midrule[0.5pt]
Original & 11.72 & 10.46 & 11.72 & 13.39 & 12.55 & 11.30 & 14.64 & 13.81 & 12.45 \\
\midrule[0.5pt]
+CodeT5 & 14.64 & 10.04 & 10.46 & 12.13 & 12.55 & 12.13 & 15.90 & 13.81 & 12.71(+0.26) \\
FEM & 12.97 & 10.88 & 11.72 & 12.55 & 12.55 & 11.72 & 15.06 & 15.06 & 12.81(+0.36) \\
\midrule[0.5pt]
+UniX & 12.13 & 9.62 & 12.13 & 12.13 & 13.39 & 11.72 & 12.97 & 12.55 & 12.08(-0.37) \\
FEM & 13.39 & 9.21 & 13.39 & 12.97 & 13.81 & 10.46 & 15.48 & 14.23 & 12.87(+0.42) \\
\midrule[0.5pt]
+BOTH & 11.72 & 10.88 & 10.04 & 15.48 & 10.46 & 11.30 & 15.48 & 14.64 & 12.50(+0.05) \\
\bottomrule[0.8pt]
\end{tabular}
\end{table}

\begin{table}[htb!]
\centering
\scriptsize 
\caption{ASR (\%) Results of CodeT5 with Different Exploration Rates and Strategies on Defeat Detection}
\setlength{\tabcolsep}{3pt}
\label{tab:codet5_results}
\begin{tabular}{lcccccccccc}
\toprule[0.8pt]
Strategy & 14\% & 21\% & 28\% & 35\% & 42\% & 49\% & 56\% & 63\% & Avg\\
\midrule[0.5pt]
Original & 41.48 & 40.64 & 40.64 & 40.64 & 42.25 & 41.18 & 42.78 & 43.85 & 41.68 \\
\midrule[0.5pt]
+CodeBE & 41.18 & 40.64 & 40.64 & 40.64 & 40.64 & 38.50 & 40.64 & 44.39 & 40.91(-0.77) \\
FEM & 33.69 & 34.22 & 33.69 & 33.69 & 32.09 & 32.62 & 35.29 & 35.29 & 33.82(-7.86) \\
\midrule[0.5pt]
+UniX & 39.57 & 37.97 & 37.97 & 22.78 & 42.25 & 37.97 & 37.97 & 42.25 & 37.34(-4.34) \\
FEM & 23.53 & 26.74 & 27.27 & 27.27 & 28.88 & 29.95 & 36.36 & 37.97 & 29.75(-11.93) \\
\midrule[0.5pt]
+BOTH & 43.85 & 40.11 & 40.11 & 40.11 & 41.18 & 40.11 & 45.99 & 45.99 & 42.18(+0.50) \\
\bottomrule[0.8pt]
\end{tabular}
\end{table}

\begin{table}[htb!]
\centering
\scriptsize
\caption{ASR (\%) Results of UniXcoder with Different Exploration Rates and Strategies on Defeat Detection}
\setlength{\tabcolsep}{3pt}
\label{tab:unixcoder_results}
\begin{tabular}{lcccccccccc}
\toprule[0.8pt]
Strategy & 14\% & 21\% & 28\% & 35\% & 42\% & 49\% & 56\% & 63\% & Avg \\
\midrule[0.5pt]
Original & 21.52 & 20.25 & 21.52 & 20.89 & 20.89 & 19.62 & 19.62 & 20.25 & 20.57 \\
\midrule[0.5pt]
+CodeBE & 18.35 & 22.78 & 22.15 & 21.52 & 18.99 & 17.09 & 18.99 & 19.62 & 20.06(-0.51) \\
FEM & 21.52 & 22.78 & 22.78 & 20.89 & 19.62 & 17.72 & 21.52 & 20.89 & 20.97(+0.40) \\
\midrule[0.5pt]
+CodeT5 & 18.99 & 21.52 & 22.78 & 18.35 & 18.35 & 15.82 & 17.09 & 20.25 & 19.14(-1.43) \\
FEM & 17.09 & 19.62 & 24.05 & 19.62 & 18.35 & 18.35 & 17.72 & 19.62 & 19.30(-1.27) \\
\midrule[0.5pt]
+BOTH & 21.52 & 19.62 & 20.89 & 18.35 & 20.25 & 18.35 & 19.62 & 19.62 & 19.94(-0.63) \\
\bottomrule[0.8pt]
\end{tabular}
\end{table}

\subsubsection{Result Analysis} 
\mbox{}\\
\textbf{Comparison With Baseline:} 
Our PATD method demonstrates substantial improvements over both baselines across all models and tasks. For Authorship Attribution, PATD achieves average improvements of 15.09\% over CodeTAE and 9.04\% over random baseline, with consistent gains across CodeBERT (14.29\%), CodeT5 (16.67\%), and UniXCoder (14.29\%). On Defeat Detection, PATD shows even more pronounced advantages with 13.0\% and 11.03\% improvements respectively, with CodeT5 achieving the most significant improvement of 22.99\% over CodeTAE.

\textbf{PGSA vs. MMMT Performance:}
Our dual auxiliary model strategy (MMMT) consistently outperforms single auxiliary model approach (PGSA) across different exploration rates. CodeBERT with UniXcoder achieves superior performance with peak ASR of 35-36\%, while CodeT5 benefits from CodeT5 and UniXcoder pairing, reaching approximately 32\% ASR. UniXcoder shows the most dramatic improvement with MMMT, achieving 34-35\% ASR at optimal exploration rates. MMMT also exhibits more stable performance across varying exploration rates.

\textbf{Full Experience Memory (FEM) Analysis:}
For Defeat Detection, FEM implementation demonstrates the value of leveraging complete experience over success-only memory. CodeBERT shows consistent modest improvements with CodeT5 (+0.36\%) and UniXcoder (+0.42\%) knowledge transfers, while UniXcoder achieves positive gains with CodeBERT knowledge (+0.40\%). These findings highlight that incorporating both successful and failed attack experiences enhances transferability compared to using only successful patterns.
\subsubsection{Discussions}
\mbox{}\\
\textbf{Memorization Pattern Complementarity:} Single-guide strategies often outperform dual-guide approaches for Authorship Attribution, indicating that memorization patterns across SCMs are not always complementary. When models provide conflicting signals, the adaptive mechanism struggles to synthesize information effectively.

\textbf{Exploration-Exploitation Balance:} Performance stability across varying exploration rates validates our hybrid approach with 30\% random exploration, preventing over-dependence on auxiliary model insights while maintaining discovery of novel vulnerabilities.

\textbf{Security Implications:} Cross-model knowledge transfer reveals that memorization patterns from one model can enhance attacks against others. Model providers should evaluate vulnerabilities within the broader ecosystem, accounting for cross-model information leakage.

\textbf{Adaptive Strategy Refinement:} Mixed results from multi-model guidance suggest naive combination of knowledge sources may introduce interference. Future work should explore sophisticated fusion mechanisms like attention-based guidance weighting or dynamic model selection.

\begin{table}[htb!]
\centering
\scriptsize 
\setlength{\tabcolsep}{2pt}
\caption{ASR (\%) Comparison on Authorship Attribution Across Different LLM4Code}
\label{tab:generative_asr_performanceaa}
\begin{tabular}{c|ccc|ccc|ccc}
\toprule[0.8pt]
Source SCMs & \multicolumn{3}{c|}{UniXcoder} & \multicolumn{3}{c|}{CodeBERT} & \multicolumn{3}{c}{CodeT5} \\
 & ASR & Recall & F1 & ASR & Recall & F1 & ASR & Recall & F1 \\
\midrule[0.5pt]
CodeLlama-7B & 28.41 & 71.59 & 83.44 & 16.84 & 83.16 & 90.80 & 41.58 & 58.42 & 73.75 \\
\midrule[0.5pt]
StarCoder & 6.72 & 93.18 & 96.47 & 9.47 & 90.53 & 95.03 & 9.90 & 91.00 & 95.29 \\
\midrule[0.5pt]
StarCoder2-3B & 7.95 & 92.05 & 95.86 & 15.79 & 84.21 & 91.43 & 6.93 & 93.07 & 96.41 \\
\bottomrule[0.8pt]
\end{tabular}
\end{table}
\begin{table}[htb!]
\centering
\scriptsize 
\setlength{\tabcolsep}{2pt}
\caption{ASR (\%) Comparison on Defeat Detection Across Different LLM4Code}
\label{tab:generative_asr_performancede}
\begin{tabular}{c|ccc|ccc|ccc}
\toprule[0.8pt]
 Source SCMs & \multicolumn{3}{c|}{UniXcoder} & \multicolumn{3}{c|}{CodeBERT} & \multicolumn{3}{c}{CodeT5} \\
 & ASR & Recall & F1 & ASR & Recall & F1 & ASR & Recall & F1 \\
\midrule[0.5pt]
CodeLlama-7B & 64.56 & 35.44 & 52.34 & 57.74 & 42.26 & 59.41 & 55.61 & 44.39 & 61.48 \\
\midrule[0.5pt]
StarCoder & 3.16 & 96.84 & 98.39 & 5.86 & 94.14 & 96.98 & 9.09 & 90.91 & 95.51 \\
\midrule[0.5pt]
StarCoder2-3B & 15.82 & 84.18 & 91.41 & 23.43 & 76.57 & 86.73 & 25.67 & 74.33 & 85.28 \\
\bottomrule[0.8pt]
\end{tabular}
\end{table}

\subsection{Evaluating on LLM4Code}

\subsubsection{Preparation}
We evaluate transferability on three representative LLM4Code: CodeLlama-7B~\cite{roziere2023code}, StarCoder\cite{li2023starcoder}, and StarCoder2-3B~\cite{lozhkov2024starcoder}. CodeLlama-7B specializes in code understanding and generation with strong reasoning capabilities, StarCoder provides robust code completion and generation across diverse programming languages, while StarCoder2-3B offers efficient code processing with broad language support.

After completing PGSA and MMMT, the victim SCM collects all ASR results, selects the best-performing parameters, and generates a JSON file containing both original and adversarial code. This file is then used to queries the LLM4Code above, instructing it to determine if two code snippets are functionally equivalent while ignoring non-executed code (like if false), unused functions/variables, and comments.

We calculate key performance metrics:
\begin{align}
\text{Precision: } \mathcal{P}_{rec} &= {TP}/{(TP + FP)} \\
\text{ASR: } \mathcal{ASR}  &= 1- \mathcal{P}_{rec} \\
\text{Recall: } \mathcal{R} &= {TP}/{(TP + FN)} \\
\text{F1-Score: } \mathcal{F}_1 &= 2 \cdot {(\mathcal{P}_{rec} \cdot \mathcal{R})}/{(\mathcal{P}_{rec} + \mathcal{R})}
\end{align}
\subsubsection{Result and Analysis}
As shown in Tab~\ref{tab:generative_asr_performanceaa} and Tab~\ref{tab:generative_asr_performancede}, our evaluation reveals distinctive vulnerability patterns across LLM4Code models. CodeLlama-7B shows the highest susceptibility with ASRs reaching 41.58\% on Authorship Attribution and 64.56\% on Defeat Detection, suggesting significant transferability from traditional SCMs. StarCoder demonstrates exceptional robustness, maintaining low ASRs (3.16\%-9.90\%) and high F1 scores (95.03\%-98.39\%) across both datasets. StarCoder2-3B exhibits moderate vulnerability, showing higher ASRs on Defeat Detection (15.82\%-25.67\%) compared to Authorship Attribution (6.93\%-15.79\%). These findings highlight that adversarial transferability varies substantially based on model architecture and dataset characteristics.

\subsubsection{Discussions}
Our study reveals significant variation in adversarial robustness among LLM4Code models. CodeLlama-7B's high vulnerability contrasts with StarCoder's exceptional resilience, suggesting fundamental architectural differences that impact security. Higher attack transferability on Defeat Detection versus Authorship Attribution indicates that task characteristics substantially influence vulnerability, suggesting security evaluations should prioritize task-specific testing. These findings highlight a critical security-capability tradeoff: as models scale to enhance performance, they may simultaneously increase vulnerability to adversarial attacks. Future work should investigate architectural elements contributing to StarCoder's robustness and develop transferability-aware training techniques.

\section{Conclusion}
This work presents the first systematic investigation of transferable adversarial vulnerabilities between traditional SCMs and LLM4Code systems. Our HABITAT method achieves up to 64\% attack success rates against LLM4Code using only traditional SCM guidance, significantly outperforming existing methods. Through comprehensive analysis of five factors governing cross-architecture vulnerability transfer, we reveal that fundamental security weaknesses persist across the entire SCM ecosystem despite architectural diversity, enabling cascading security failures. Our findings provide essential insights for developing robust adversarial defenses and proactively strengthening SCM systems against cross-architecture attacks.

\bibliographystyle{elsarticle-harv} 
\bibliography{example}






\end{document}